\documentstyle[aps,12pt,preprint,psfig]{revtex}
\tightenlines

\begin{document}
\preprint{
\hbox{PKU-TP-98-53}}
\draft

\title{Diffractive light quark jet production at
        hadron colliders in the two-gluon exchange model}

\author{Feng Yuan}
\address{\small {\it Department of Physics, Peking University, Beijing 100871, People's Republic
of China}}
\author{Kuang-Ta Chao}
\address{\small {\it China Center of Advanced Science and Technology (World Laboratory), Beijing 100080,
        People's Republic of China\\
      and Department of Physics, Peking University, Beijing 100871, People's Republic of China}}

\maketitle
\begin{abstract}

Massless quark and antiquark jet production at large transverse momentum in
the coherent diffractive processes at hadron colliders is calculated in the
two-gluon exchange parametrization of the Pomeron model. We use the helicity
amplitude method to calculate the cross section formula. We find that for the
light quark jet production the diffractive process is related to the
differential off-diagonal gluon distribution function in the proton.
We estimate the production rate for this process at the Fermilab Tevatron
by approximating the off-diagonal gluon distribution function by the usual
diagonal gluon distribution in the proton. And we find that the cross
sections for the diffractive light quark jet production and the charm
quark jet production are in the same order of magnitude.
We also use the helicity amplitude method to calculate the diffractive
charm jet production at hadron colliders, by which we reproduce the leading
logarithmic approximation result of this process we previously calculated.
\end{abstract}


\section{Introduction}

In recent years, there has been a renaissance of interest in
diffractive scattering.
These diffractive processes are described by the Regge theory in
terms of the Pomeron ($I\!\! P$) exchange\cite{pomeron}.
The Pomeron carries quantum numbers of the vacuum, so it is a colorless entity
in QCD language, which may lead to the ``rapidity gap" events in experiments.
However, the nature of Pomeron and its reaction with hadrons remain a mystery.
For a long time it had been understood that the dynamics of the
``soft pomeron'' is deeply tied to confinement.
However, it has been realized now that how much can be learned about
QCD from the wide variety of small-$x$ and hard diffractive processes,
which are now under study experimentally.
In Refs.\cite{th1,th2}, the diffractive $J/\psi$ and $\Upsilon$ production
cross section have been formulated in photoproduction processes and in
DIS processes in perturbative QCD.
In the framework of perturbative QCD the Pomeron is represented by a pair of
gluon in the color-singlet sate.
This two-gluon exchange model can successfully describe the experimental
results from HERA\cite{hera-ex}.

On the other hand, as we know that there exist nonfactorization effects
in the hard diffractive processes at hadron colliders
\cite{preqcd,collins,soper,tev}.
First, there is the so-called spectator effect\cite{soper}, which can
change the probability of the diffractive hadron emerging from collisions
intact. Practically, a suppression factor (or survive factor) ``$S_F$"
is used to describe this effect\cite{survive}.
Obviously, this suppression factor can not be calculated in perturbative
QCD, which is now viewed as a nonperturbative parameter.
Typically, the suppression factor $S_F$ is determined to be about
$0.1$ at the energy scale of the Fermilab Tevatron\cite{tev}.
Another nonfactorization effect discussed in literature is associated with the coherent
diffractive processes at hadron colliders\cite{collins}, in which
the whole Pomeron is induced in the hard scattering.
It is proved in \cite{collins} that the existence of the leading twist
coherent diffractive processes is associated with a breakdown of the
QCD factorization theorem.

Based on the success of the two-gluon exchange parametrization of the Pomeron
model in the description of the diffractive photoproduction processes
at $ep$ colliders\cite{th1,th2,hera-ex}, we may extend the applications
of this model to calculate the diffractive processes at hadron colliders in perturbative QCD.
Under this context, the Pomeron represented by a color-singlet two-gluon
system emits from one hadron and interacts with another hadron in hard
process, in which the two gluons are both involved (as shown in Fig.~1).
The partonic process is plotted in Fig.2, where there are nine diagrams
in the leading order of perturbative QCD.
Therefore, these processes calculated in the two-gluon exchange model are just
belong to the coherent diffractive processes in hadron collisions.
Another important feature of the calculations of the diffractive processes in this model recently demonstrated is the sensitivity
to the off-diagonal parton distribution
function in the proton\cite{offd}.

Using this two-gluon exchange model, we have calculated the  diffractive
$J/\psi$ production \cite{psi}, charm jet production\cite{charm},
massive muon pair and $W$ boson productions\cite{dy} in hadron collisions.
These calculations show that we can explore much low $x$ (off-diagonal)
gluon distribution function and study the coherent diffractive processes
at hadron colliders through these processes.
In this paper, we will calculate the light quark (massless) jet production
at large transverse momentum in the coherent diffractive processes at hadron
colliders by using the two-gluon exchange model.
In the calculations of Refs.\cite{psi,charm,dy}, there always is a large mass
scale associated with the production process.
That is $M_\psi$ for $J/\psi$ production, $m_c$ for the charm jet
production, $M^2$ for the massive muon production ($M^2$ is the invariant mass
of the muon pair) and $M_W^2$ for $W$ boson production.
However, in the light quark jet production process, 
there is no large mass scale.
So, for the light quark jet production, the large transverse momentum is needed to
guarantee the application of the perturbative QCD.
Furthermore, the experience on the calculations of the diffractive di-quark
photoproduction\cite{zaka} shows that the light quark jet production
in the two-gluon exchange model has a distinctive feature that
there is no contribution from the small $l_T^2$ region ($l_T^2<k_T^2$)
in the integration of the amplitude over $l_T^2$.
So, the expansion (in terms of $l_T^2/M_X^2$) method 
used in Refs.\cite{psi,charm,dy} can not be applied to the calculations
of light quark jet production.
In the following calculations, we will employ the helicity amplitude method
to calculate the amplitude of the diffractive light quark jet production
in hadron collisions.
We will show that the production
cross section is related to the differential (off-diagonal) gluon distribution
function in the proton as that in the diffractive di-quark jet
photoprodution process\cite{zaka}.
(On the other hand, we note that the cross sections of the processes
calculated in Refs.\cite{psi,charm,dy} are related to 
the integrated gluon distribution function in the proton).

The diffractive production of heavy quark jet at hadron colliders has also
been studied by using the two-gluon exchange model in Ref.\cite{levin}.
However, their calculation method is very different from ours
\footnote{For detailed discussions and comments, please see \cite{charm}}.
In their calculations, they separated their diagrams into two parts,
and called one part the coherent diffractive contribution to the heavy
quark production.
However, this separation can not guarantee the
gauge invariance\cite{charm}.
In our approach, we follow the definition of
Ref.\cite{collins}, i.e., we call
the process in which the whole Pomeron participants in the hard scattering
process as the coherent diffractive process.
Under this definition, all of the diagrams plotted in Fig.2
for the partonic process $gp\rightarrow q\bar qp$ contribute to
the coherent diffractive production.

The rest of the paper is organized as follows.
In Sec.II, we will give the cross section formula for the partonic process
$gp\rightarrow q\bar q p$ in the leading order of perturbative QCD,
where we employ the helicity amplitude method to calculate the amplitude
for this process.
In Sec.III, we use this helicity amplitude method to recalculate the diffractive
charm jet production process $gp\rightarrow c\bar c p$, by which
we will reproduce the leading logarithmic approximation result for this
process previously calculated in Ref.\cite{charm}.
In Sec.IV, we will estimate the production rate of diffractive light
quark jet at the Fermilab Tevatron by approximating the off-diagonal
gluon distribution function by the usual diagonal gluon distribution
function in the proton.
And the conclusions will be given in Sec.V.

\section{ The cross section formula for the partonic process}

For the partonic process $gp\rightarrow q\bar qp$, in the leading order of perturbative
QCD, there are nine diagrams shown in Fig.2.
The two-gluon system coupled to the proton (antiproton) in Fig.2 is in
a color-singlet state, which characterizes the diffractive processes in
perturbative QCD.
Due to the positive signature of these diagrams (color-singlet exchange),
we know that the real part of the amplitude cancels out in the leading
logarithmic approximation.
To get the imaginary part of the amplitude, we must calculate the
discontinuity represented by the crosses in each diagram of Fig.2.

The first four diagrams of Fig.2 are the same as those calculated in the
diffractive photoproduction processes. But, due to the existence of
gluon-gluon interaction vertex in QCD, in the partonic process
$gp\rightarrow q\bar q p$, there are additional five diagrams (Fig.2(5)-(9)).
These five diagrams are needed for complete calculations in this order of QCD.

In our calculations, we express the formulas in terms of the Sudakov
variables.
That is, every four-momenta $k_i$ are decomposed as,
\begin{equation}
k_i=\alpha_i q+\beta_i p+\vec{k}_{iT},
\end{equation}
where $q$ and $p$ are the momenta of the incident gluon and the proton,
$q^2=0$, $p^2=0$, and $2p\cdot q=W^2=s$.
Here $s$ is the c.m. energy of the gluon-proton system, i.e., the invariant
mass of the partonic process $gp\rightarrow q\bar q p$.
$\alpha_i$ and $\beta_i$ are the momentum fractions of $q$ and $p$
respectively.
$k_{iT}$ is the transverse momentum, which satisfies
\begin{equation}
k_{iT}\cdot q=0,~~~
k_{iT}\cdot p=0.
\end{equation}

All of the Sudakov variables for every momentum
are determined by using the on-shell conditions
of the momenta represented by the external lines and the crossed lines in the diagram.
The calculations of these Sudakov variables are similar to those in
the diffractive charm jet production process $gp\rightarrow c\bar c p$.
And all of these Sudakov variables for the process $gp\rightarrow q\bar qp$
of the light quark jet production can take their values from the corresponding
formulas in Ref.\cite{charm} after taking $m_c\rightarrow 0$.
For convenience, we list all of the Sudakov variables for the diffractive
process $gp\rightarrow q\bar qp$ in the following.

For the momentum $u$, we have
\begin{equation}
\alpha_u=0,~~\beta_u=x_{I\! P}=\frac{M_X^2}{s},~~u_T^2=t=0,
\end{equation}
where $M_X^2$ is the invariant mass squared of the diffractive final state
including the light quark and antiquark jets.
For the high energy diffractive process, we know that $M_x^2\ll s$, so
we have $\beta _u$ ($x_{I\! P}$) as a small parameter.
For the momentum $k$,
\begin{equation}
\label{ak}
\alpha_k(1+\alpha_k)=-\frac{k_T^2}{M_X^2},~~\beta_k=-\alpha_k\beta_u,
\end{equation}
where $k_T$ is the transverse momentum of the out going quark jet.
For the loop mentum $l$,
\begin{eqnarray}
\nonumber
\alpha_l&=&-\frac{l_T^2}{s},\\
\nonumber
\beta_l&=&\frac{2(k_T,l_T)-l_T^2}{\alpha_ks},~~~{\rm for~Diag.}1,~3,~5,\\
\nonumber
&=&\frac{2(k_T,l_T)+l_T^2}{(1+\alpha_k)s},~~~{\rm for~Diag.}2,~4,~6,\\
&=&-\frac{M_X^2-l_T^2}{s},~~~~~~~{\rm for~Diag.}7,~8,~9.
\end{eqnarray}

Using these Sudakov variables, we can give the cross section formula
for the partonic process $gp\rightarrow q\bar q p$ as,
\begin{equation}
\label{xs}
\frac{d\hat{\sigma}(gp\rightarrow q\bar qp)}{dt}|_{t=0}=\frac{dM_X^2d^2k_Td\alpha_k}{16\pi s^216\pi^3M_X^2}
        \delta(\alpha_k(1+\alpha_k)+\frac{k_T^2}{M_X^2})\sum \overline{|{\cal A}|}^2,
\end{equation}
where ${\cal A}$ is the amplitude of the process $gp\rightarrow q\bar qp$.
We know that the real part of the amplitude ${\cal A}$ is zero,
and the imaginary part of the amplitude ${\cal A}(gp\rightarrow q\bar qp)$ for each diagram
of Fig.2 has the following general form,
\begin{equation}
\label{ima}
{\rm Im}{\cal A}=C_F(T_{ij}^a)\int \frac{d^2l_T}{(l_T^2)^2}F\times\bar u
        _i(k+q)\Gamma_\mu v_j(u-k),
\end{equation}
where $C_F$ is the color factor for each diagram, and is the same as that
for the $gp\rightarrow c\bar c p$ process\cite{charm}.
$a$ is the color index of the incident gluon.
$\Gamma_\mu$ represents some $\gamma$ matrices including one
propagator. $F$ in the integral is defined as
\begin{equation}
F=\frac{3}{2s}g_s^3f(x',x^{\prime\prime};l_T^2),
\end{equation}
where
\begin{equation}
\label{offd1}
f(x',x^{\prime\prime};l_T^2)=\frac{\partial G(x',x^{\prime\prime};l_T^2)}{\partial {\rm ln} l_T^2},
\end{equation}
where the function
$G(x',x^{\prime\prime};k_T^2)$ is the so-called
off-diagonal gluon distribution function\cite{offd}.
Here, $x'$ and $x^{\prime\prime}$ are the momentum fractions of the proton
carried by the two gluons.
It is expected that at small $x$, there is no big difference between the off-diagonal and
the usual diagonal gluon densities\cite{off-diag}.
So, in the following calculations, we estimate the production rate by
approximating the off-diagonal gluon density by 
the usual diagonal gluon density, 
$G(x',x^{\prime\prime};Q^2)\approx xg(x,Q^2)$, where $x=x_{I\!\! P}=M_X^2/s$.

In Ref.\cite{charm}, we calculate the amplitude (\ref{ima}) for the heavy quark
diffractive production processes by expanding
$\Gamma_\mu$ in terms of $l_T^2$. However, in the light quark jet production
process $gp\rightarrow q\bar qp$, the expansion method is not yet valid.
According to the result of \cite{charm}, the production cross section
is proportional to the heavy quark mass. If we apply this formula to the
light quark jet production, the cross section will be zero.
That is to say, the expansion of the amplitude in terms of $l_T^2$, in which
the large logarithmic contribution comes from the region of $l_T^2\ll M_X^2$,
is not further suitable for the calculation of the cross section for massless light
quark jet production.
Furthermore, the experience of the calculation of the diffractive light quark
jet photoproduction process $\gamma p\rightarrow q\bar qp$ (for $Q^2=0$)
\cite{zaka} indicates that there is no contribution from
the region of $l_T^2< k_T^2$ in the integration of the amplitude over
$l_T^2$.
In the hadroproduction process $gp\rightarrow q\bar qp$,
the situation is the same.
So, the expansion method, in which $l_T^2$ is taken as a small
parameter, is not valid for the calculations of the massless quark production
processes.

In the following, we employ the helicity amplitude method\cite{ham} to calculate
the amplitude Eq.(\ref{ima}). For the massless quark spinors, we define
\begin{equation}
u_\pm(p)=\frac{1}{\sqrt{2}}(1\pm\gamma_5)u(p).
\end{equation}
For the polarization vector of the incident gluon, which is transversely polarized,
we choose,
\begin{equation}
\label{ev}
e_\pm=\frac{1}{\sqrt{2}}(0,1,\pm i,0).
\end{equation}
The helicity amplitudes for the processes in which the polarized Dirac particles are involved
have the following general forms\cite{ham},
\begin{equation}
\label{ham}
\bar u_\pm(p_f)Qv_\mp(p_i)=\frac{Tr[Q\not\! p_i\not\! n\not\! p_f(1\mp\gamma_5)]}
        {4\sqrt{(n\cdot p_i)(n\cdot p_f)}},
\end{equation}
where $n$ is an arbitrary massless 4-vector, which is set to be $n=p$ in the
following calculations.
Using this formula (\ref{ham}), the calculations of the helicity amplitude
${\cal A}(\lambda(g),\lambda(q),\lambda({\bar q}))$ for the diffractive
process $gp\rightarrow q\bar q p$ is straightforward.
Here $\lambda$ are the corresponding helicities of the external gluon,
quark and antiquark.
In our calculations, we only take the leading order contribution, and neglect the
higher order contributions which are proportional to $\beta_u=\frac{M_X^2}{s}$
because in the high energy diffractive processes we have $\beta_u\ll 1$.

For the first four diagrams, to sum up together, the imaginary part of the
amplitude ${\cal A}(\pm,+,-)$ is
\begin{equation}
\label{im1}
{\rm Im}{\cal A}^{1234}(\pm,+,-)=\alpha_k^2(1+\alpha_k){\cal N}\times
        \int\frac{d^2\vec{l}_T}{(l_T^2)^2}f(x',x'';l_T^2)
        (-\frac{2}{9}\frac{\vec{e}^{(\pm)}\cdot \vec{k}_T}{k_T^2}-\frac{1}{36}\frac{\vec{e}^{(\pm)}\cdot(\vec{k}_T-\vec{l}_T)}{(\vec{k}_T-\vec{l}_T)^2}),
\end{equation}
where $\frac{2}{9}$ and $-\frac{1}{36}$ are the color factors for Diag.1,4 and
Diag.2,3 respectively, and
${\cal N}$ is defined as
\begin{equation}
\label{ne}
{\cal N}=\frac{s}{\sqrt{-\alpha_k(1+\alpha_k)}}g_s^3T_{ij}^a.
\end{equation}
The other helicity amplitudes for the first four diagrams have the similar form
as (\ref{im1}).
The amplitude expression Eq.~(\ref{im1}) is the same as that for the photoproduction
process $\gamma p\rightarrow q\bar q p$ previously calculated in Refs.\cite{th3,zaka} except the difference
on the color factors. In the diffractive photoproduction process, the color
factors of these four diagrams are the same (they are all $\frac{2}{9}$), while in hadroproduction
process the color factors are no longer the same for these four diagrams.
It is instructive to see what is the consequence of this difference.
We know that the amplitude of the diffractive process in Eq.~(\ref{ima}) must be
zero in the limit $l_T^2\rightarrow 0$. Otherwise, this will lead to a linear singularity
when we perform the integration of the amplitude over $l_T^2$ due to existence of the factor $1/(l_T^2)^2$ in the
integral of Eq.~(\ref{ima})\cite{charm}.
This linear singularity is not proper in QCD calculations.
So, we must first exam the amplitude behavior under the limit of $l_T^2\rightarrow 0$
for all the diffractive processes in the calculations using the two-gluon exchange model.
From Eq.~(\ref{im1}), we can see that the amplitude for the
diffractive photoproduction of di-quark jet process is exact zero at $l_T^2\rightarrow 0$.
However, for the diffractive hadroproduction process $gp\rightarrow q\bar qp$ the amplitude
for the first four diagrams 
is not exact zero in the limit $l_T^2\rightarrow 0$ due to the inequality of the
color factors between them.
So, for $gp\rightarrow q\bar qp$ process there must be other diagrams in this order
of perturbative QCD calculation to cancel out the linear singularity which
rises from the first four diagrams.
The last five diagrams of Fig.2 are just for this purpose.

For example, the contributions from Diags.5 and 8 are
\begin{equation}
{\rm Im}{\cal A}^{58}(\pm,+,-)=\alpha_k^2(1+\alpha_k){\cal N}\times
        \int\frac{d^2\vec{l}_T}{(l_T^2)^2}f(x',x'';l_T^2)
        (-\frac{1+\alpha_k}{4}\frac{\vec{e}^{(\pm)}\cdot (\vec{k}_T-(1+\alpha_k)\vec{l}_T)}{(\vec{k}_T-(1+\alpha_k)^2\vec{l}_T)^2}),
\end{equation}
and the contributions from Diags.6 and 9 are 
\begin{equation}
{\rm Im}{\cal A}^{69}(\pm,+,-)=\alpha_k^2(1+\alpha_k){\cal N}\times
        \int\frac{d^2\vec{l}_T}{(l_T^2)^2}f(x',x'';l_T^2)
        (\frac{\alpha_k}{4}\frac{\vec{e}^{(\pm)}\cdot (\vec{k}_T-\alpha_k\vec{l}_T)}{(\vec{k}_T-\alpha_k^2\vec{l}_T)^2}),
\end{equation}
and the contribution from Diag.7 is
\begin{equation}
{\rm Im}{\cal A}^{7}(\pm,+,-)=\alpha_k^2(1+\alpha_k){\cal N}\times
        \int\frac{d^2\vec{l}_T}{(l_T^2)^2}f(x',x'';l_T^2)
        (\frac{1}{2}\frac{\vec{e}^{(\pm)}\cdot \vec{k}_T}{k_T^2}).
\end{equation}
From the above results, we can see that the contributions from Diags.5-9
just cancel out the linear singularity which rises from the first four diagrams.
Their total sum from the nine diagrams of Fig.2 is free of linear
singularity now.

Finally, by adding up all of the nine diagrams of Fig.2, the
imaginary parts of the amplitudes for the following helicity sets
are,
\begin{eqnarray}
\nonumber
{\rm Im}{\cal A}(\pm,+,-)&=&\alpha_k^2(1+\alpha_k){\cal N}\times {\cal T}^{(\pm)},\\
{\rm Im}{\cal A}(\pm,-,+)&=&\alpha_k(1+\alpha_k)^2{\cal N}\times {\cal T}^{(\pm)},
\end{eqnarray}
where
\begin{eqnarray}
\label{int2}
\nonumber
{\cal T}^{(\pm)}&=&\int\frac{d^2\vec{l}_T}{(l_T^2)^2}f(x',x'';l_T^2)[(\frac{1}{2}-\frac{2}{9})
        \frac{\vec{e}^{(\pm)}\cdot \vec{k}_T}{k_T^2}-\frac{1}{36}\frac{\vec{e}^{(\pm)}\cdot(\vec{k}_T-\vec{l}_T)}{(\vec{k}_T-\vec{l}_T)^2}\\
&&-\frac{1+\alpha_k}{4}\frac{\vec{e}^{(\pm)}\cdot(\vec{k}_T-(1+\alpha_k)\vec{l}_T)}{(\vec{k}_T-(1+\alpha_k)\vec{l}_T)^2}
  +\frac{\alpha_k}{4}\frac{\vec{e}^{(\pm)}\cdot(\vec{k}_T-\alpha_k\vec{l}_T)}{(\vec{k}_T-\alpha_k\vec{l}_T)^2}].
\end{eqnarray}
And the amplitudes for the other helicity sets are zero in the light quark jet
production process.
From the above results, we can see that in the integration of the amplitude
the linear singularity from different diagrams are canceled out by each other,
which will guarantee there is no linear singularity in the total sum.

Another feature of the above results for the amplitudes is the relation to the
differential off-diagonal gluon distribution function $f(x',x'';l_T^2)$.
To get the cross section for the diffractive process, we must perform the
integration of Eq.~(\ref{int2}).
However, as mentioned above that there is no big difference between the off-diagonal
gluon distribution function and the usual gluon distribution at small $x$,
so we can simplify the integration of (\ref{int2}) by 
approximating the differential off-diagonal gluon
distribution function $f(x',x'';l_T^2)$ by the usual diagonal differential
gluon distribution function $f_g(x;l_T^2)$.

After integrating over the azimuth angle of $\vec{l}_T$, the integration
${\cal T}^{(\pm)}$ will then be
\begin{equation}
{\cal T}^{(\pm)}=\pi\frac{\vec{e}^{(\pm)}\cdot \vec{k}_T}{k_T^2}{\cal I},
\end{equation}
where
\begin{eqnarray}
\label{inti}
\nonumber
{\cal I}&=&\int\frac{dl_T^2}{(l_T^2)^2}f_g(x;l_T^2)[\frac{1}{36}(\frac{1}{2}-\frac{k_T^2-l_T^2}{2|k_T^2-l_T^2|})
        +\frac{1+\alpha_k}{4}(\frac{1}{2}-\frac{k_T^2-(1+\alpha_k)l_T^2}{2|k_T^2-(1+\alpha_k)l_T^2|})\\
        &&-\frac{\alpha_k}{4}(\frac{1}{2}-\frac{k_T^2-\alpha_k l_T^2}{2|k_T^2-\alpha_k l_T^2|})].
\end{eqnarray}
Comparing the above results with those of the photoproduction process $\gamma p\rightarrow q\bar q p$\cite{th3,zaka},
we find that the amplitude formula for the diffractive light quark jet hadroproduction 
process $gp\rightarrow q\bar qp$ is much more complicated.
However, the basic structure of the amplitude, especially the expression for the
integration ${\cal I}$ is similar to that for the photoproduction process.
In the integration of (\ref{inti}), if $l_t^2<k_T^2$ the first term of the
integration over $l_T^2$ will be zero; if $l_t^2<k_T^2/(1+\alpha_k)^2$ the second
term will be zero; if $l_t^2<k_T^2/\alpha_k^2$ the third term will be zero.
So, the dominant regions contributing to the three integration terms are
$l_t^2\sim k_T^2$, $l_t^2\sim k_T^2/(1+\alpha_k)^2$, and $l_t^2\sim k_T^2/\alpha_k^2$
respectively.
Approximately, by ignoring some evolution effects of the differential gluon
distribution function $f_g(x;l_T^2)$ in the above dominant integration regions,
we get the following results for the integration ${\cal I}$,
\begin{equation}
\label{i1}
{\cal I}=\frac{1}{k_T^2}[\frac{1}{36}f_g(x;k_T^2)+\frac{(1+\alpha_k)^3}{4}f_g(x;\frac{k_T^2}{(1+\alpha_k)^2})
        -\frac{\alpha_k^3}{4}f_g(x;\frac{k_T^2}{\alpha_k^2})].
\end{equation}

Obtained the formula for the integration ${\cal I}$, the amplitude squared for
the partonic process $gp\rightarrow q\bar qp$ will be reduced to, after averaging
over the spin and color degrees of freedom,
\begin{equation}
\overline{|{\cal A}|}^2=\frac{9}{4}\alpha_s^3(4\pi)^3\pi^2s^2\frac{|{\cal I}|^2}{M_X^2}
        (1-\frac{2k_T^2}{M_X^2}).
\end{equation}
And the cross section for the partonic process $gp\rightarrow q\bar qp$ is
\begin{equation}
\label{xsp}
\frac{d\hat\sigma(gp\rightarrow q\bar qp)}{dt}|_{t=0}=\int_{M_X^4>4k_T^2}dM_X^2dk_T^2
        \frac{9\alpha_s^3\pi^2}{8(M_X^2)^2}\frac{1}{\sqrt{1-\frac{4k_T^2}{M_X^2}}}
        (1-\frac{2k_T^2}{M_X^2})|{\cal I}|^2.
\end{equation}
The integral bound $M_X^2>4k_T^2$ above shows that the dominant contribution
of the integration over $M_X^2$ comes from the region of $M_X^2\sim 4k_T^2$.
Using Eq.~(\ref{ak}), this indicates that in this dominant region 
$\alpha_k$ is of order of 1.
So, in the integration ${\cal I}$ the differential gluon distribution
function $f_g(x;Q^2)$ of the three terms can approximately take their values
at the same scale of $Q^2=k_T^2$.
That is, the integration ${\cal I}$ is then simplified to
\begin{equation}
\label{i2}
{\cal I}\approx \frac{10M_X^2-27k_T^2}{36M_X^2}f_g(x;k_T^2).
\end{equation}
Numerical calculations show that there is little difference (within $10\%$
for $k_T>5~GeV$) between the cross sections by using these two different
parametrizations of ${\cal I}$, Eq.~(\ref{i1}) and Eq.~(\ref{i2}).
So, in Sec.IV, we use Eqs.~(\ref{xsp}) and (\ref{i2}) to
estimate the diffractive light quark jet production rate at the Fermilab
Tevatron.

\section{Recalculate the heavy quark jet production using the helicity amplitude method}

For a crossing check, in this section we will recalculate the diffractive
heavy quark jet production at hadron colliders by using the helicity amplitude
method. In Ref.\cite{charm}, we have calculated this process in the leading
logarithmic approximation of QCD, where we expanded the amplitude in terms of
$l_T^2$. Now, if we use the helicity amplitude method, we donot need to use
the expansion method for the $\Gamma_\mu$ factor in Eq.~(\ref{ima}) as in \cite{charm}.
We can firstly calculate the amplitude explicitly by using the helicity amplitude
method.

However, for the massive fermion, the amplitude formula is more complicated.
Following Ref.\cite{ham1}, we first define the basic spinors $u_\pm(k_0)$ as,
\begin{eqnarray}
\label{k1}
\nonumber
u_+(k_0)=\not\! k_1u_-(k_0),\\
u_-(k_0)\bar u_-(k_0)=\frac{1}{2}(1-\gamma_5)\not\! k_0,
\end{eqnarray}
where the momenta $k_0$ and $k_1$ satisfy the followoing relations,
\begin{equation}
\label{k2}
k_0\cdot k_0=0,~~~k_1\cdot k_1=-1,~~~k_0\cdot k_1=0.
\end{equation}
Using Eqs.~(\ref{k1}) and (\ref{k2}), we can easily find that the spinor $u_+(k_0)$
satisfies
\begin{equation}
u_+(k_0)\bar u_+(k_0)=\frac{1}{2}(1+\gamma_5)\not\! k_0.
\end{equation}
Provided the basic spinors, we then express any spinors $u(p_i)$ in terms of the basic
ones,\cite{ham1}
\begin{equation}
u_\pm(p_i)=\frac{(\not\! p_i +m_i)u_\pm(k_0)}{\sqrt{2p_i\cdot k_0}}.
\end{equation}
It is easily checked that these spinors satisfy Dirac's equations.
Now, for the massive fermions, the helicity amplitudes for the processes involving
Dirac particles have the following general forms,
\begin{eqnarray}
\label{ham1}
\nonumber
\bar u_+(p_f)Qv_-(p_i)&=&\frac{Tr[Q(\not\! p_i-m_i)(1-\gamma_5)\not\! k_0(\not\! p_f+m_f)}
        {4\sqrt{(n\cdot p_i)(n\cdot p_f)}},\\
\nonumber
\bar u_-(p_f)Qv_+(p_i)&=&\frac{Tr[Q(\not\! p_i-m_i)(1+\gamma_5)\not\! k_0(\not\! p_f+m_f)}
        {4\sqrt{(n\cdot p_i)(n\cdot p_f)}},\\
\nonumber
\bar u_+(p_f)Qv_+(p_i)&=&\frac{Tr[Q(\not\! p_i-m_i)(1-\gamma_5)\not\! k_1\not\! k_0(\not\! p_f+m_f)}
        {4\sqrt{(n\cdot p_i)(n\cdot p_f)}},\\
\bar u_-(p_f)Qv_-(p_i)&=&\frac{Tr[Q(\not\! p_i-m_i)\not\! k_1(1-\gamma_5)\not\! k_0(\not\! p_f+m_f)}
        {4\sqrt{(n\cdot p_i)(n\cdot p_f)}},
\end{eqnarray}
where $m_i$ and $m_f$ are the masses for the momenta $p_i$ and $p_f$ respectively,
where $p_i^2=m_i^2,~p_f^2=m_f^2$.
From the above equations, we can see that for the massless fermions ($m_i=m_f=0$)
the formula Eq.~(\ref{ham1}) will then turn back to the formula Eq.~(\ref{ham}).

To calculate the imaginary part of the amplitude Eq.~(\ref{ima}) for the partonic process
$gp\to c\bar c p$, a convenient choice
for the momenta $k_0$ and $k_1$ is,
\begin{equation}
k_0=p,~~~~k_1=e,
\end{equation}
where the vector $e$ is the polarization vector for the incident gluon defined
in Eq.~(\ref{ev}).
Using the formula Eq.~(\ref{ham1}) and the above choice for the momenta $k_0$ and $k_1$,
the helicity amplitudes for Eq.~(\ref{ima}) will then be,
\begin{eqnarray}
\nonumber
{\rm Im}{\cal A}(\pm,+,-)&=&\alpha_k^2(1+\alpha_k){\cal N}\times {\cal T}_c^{(\pm)},\\
{\rm Im}{\cal A}(\pm,-,+)&=&\alpha_k(1+\alpha_k)^2{\cal N}\times {\cal T}_c^{(\pm)},\\
{\rm Im}{\cal A}(\pm,+,+)&=&{\rm Im}{\cal A}(\pm,-,-)=\alpha_k(1+\alpha_k){\cal N}\times \frac{\pi m_c}{2}{\cal I}'_c,
\end{eqnarray}
where ${\cal N}$ is the same as in Eq.~(\ref{ne}), and the integrations ${\cal T}_c^{(\pm)}$ and ${\cal I}'_c$
are defined as 
\begin{eqnarray}
\label{int3}
\nonumber
{\cal T}_c^{(\pm)}&=&\int\frac{d^2\vec{l}_T}{(l_T^2)^2}f(x',x'';l_T^2)[(\frac{1}{2}-\frac{2}{9})
        \frac{\vec{e}^{(\pm)}\cdot \vec{k}_T}{k_T^2+m_c^2}-\frac{1}{36}\frac{\vec{e}^{(\pm)}\cdot(\vec{k}_T-\vec{l}_T)}{m_c^2+(\vec{k}_T-\vec{l}_T)^2}\\
&&-\frac{1+\alpha_k}{4}\frac{\vec{e}^{(\pm)}\cdot(\vec{k}_T-(1+\alpha_k)\vec{l}_T)}{m_c^2+(\vec{k}_T-(1+\alpha_k)\vec{l}_T)^2}
  +\frac{\alpha_k}{4}\frac{\vec{e}^{(\pm)}\cdot(\vec{k}_T-\alpha_k\vec{l}_T)}{m_c^2+(\vec{k}_T-\alpha_k\vec{l}_T)^2}],\\
\nonumber
{\cal I}'_c&=&\frac{1}{\pi}\int\frac{d^2\vec{l}_T}{(l_T^2)^2}f(x',x'';l_T^2)[(\frac{1}{2}-\frac{2}{9})
        \frac{1}{k_T^2+m_c^2}-\frac{1}{36}\frac{1}{m_c^2+(\vec{k}_T-\vec{l}_T)^2}\\
&&-\frac{1+\alpha_k}{4}\frac{1}{m_c^2+(\vec{k}_T-(1+\alpha_k)\vec{l}_T)^2}
  +\frac{\alpha_k}{4}\frac{1}{m_c^2+(\vec{k}_T-\alpha_k\vec{l}_T)^2}].
\end{eqnarray}
If we approximate the differential off-diagonal gluon
distribution function $f(x',x'';l_T^2)$ by the usual diagonal differential
gluon distribution function $f_g(x;l_T^2)$, the above integrations will then
be reduced to, after integrating over the azimuth angle of $\vec{l}_T$,
\begin{equation}
{\cal T}_c^{(\pm)}=\pi\vec{e}^{(\pm)}\cdot \vec{k}_T{\cal I}_c,
\end{equation}
where
\begin{eqnarray}
\label{int4}
\nonumber
{\cal I}_c&=&\int\frac{d^2\vec{l}_T}{(l_T^2)^2}f_g(x;l_T^2)[\frac{5}{18}
        \frac{1}{m_T^2}-\frac{5}{36}\frac{1}{k_T^2}-\frac{1}{72}\frac{k_T^2-m_c^2-l_T^2}{m_1^2}\\
        &&-\frac{1+\alpha_k}{8}\frac{k_T^2-m_c^2-(1+\alpha_k)^2l_T^2}{m_2^2} +\frac{\alpha_k}{8}\frac{k_T^2-m_c^2-\alpha_k^2l_T^2}{m_3^2}],
\end{eqnarray}
where
\begin{eqnarray}
\nonumber
m_T^2&=&k_T^2+m_c^2,~~m_1^2=\sqrt{(m_t^2+l_T^2)^2-k_T^2l_T^2},~~m_2^2=\sqrt{(m_t^2+(1+\alpha_k)^2l_T^2)^2-(1+\alpha_k)^2k_T^2l_T^2},\\
        ~~m_3^2&=&\sqrt{(m_t^2+\alpha_k^2l_T^2)^2-\alpha_k^2k_T^2l_T^2},
\end{eqnarray}
and
\begin{equation}
\label{int5}
{\cal I}'_c=\int\frac{d^2\vec{l}_T}{(l_T^2)^2}f_g(x;l_T^2)[\frac{5}{18}
        \frac{1}{m_T^2}-\frac{1}{36}\frac{1}{m_1^2}-
        \frac{1+\alpha_k}{4}\frac{1}{m_2^2} +\frac{\alpha_k}{4}\frac{1}{m_3^2}].
\end{equation}
So, the amplitude squared for
the partonic process $gp\rightarrow c\bar cp$ will then be reduced to, after averaging
over the spin and color degrees of freedom,
\begin{equation}
\label{cam}
\overline{|{\cal A}|}^2=\frac{9}{4}\alpha_s^3(4\pi)^3\pi^2s^2\frac{m_T^2}{M_X^2}
        [(1-\frac{2k_T^2}{M_X^2})k_T^2|{\cal I}_c|^2+m_c^2|{\cal I}'_c|^2].
\end{equation}

From the above results, we can see that the integrals of Eqs.~(\ref{int4}) and
(\ref{int5}) are proportional to $1/l_T^2$ in the limit of $l_T^2\rightarrow 0$.
That is to say that there exist large logarithmic contributions from the integration region
of $1/R^2_N\ll l_T^2\ll m_T^2$ for the integration over $l_T^2$ as in Ref.\cite{charm}.
So, we can expand the integrals of Eqs.~(\ref{int4}) and (\ref{int5}) in terms
of $l_T^2$ to get the leading logarithmic contribution to the amplitude.
In the limit of $l_T^2\rightarrow 0$, the parameters $m_1^2$, $m_2^2$ and $m_3^2$
scale as,
\begin{eqnarray}
\nonumber
\frac{1}{m_1^2}&\approx & \frac{1}{m_T^2}[1-\frac{m_c^2-k_T^2}{m_T^2}\frac{l_T^2}{m_T^2}],\\
\nonumber
\frac{1}{m_2^2}&\approx & \frac{1}{m_T^2}[1-\frac{m_c^2-k_T^2}{m_T^2}\frac{(1+\alpha_k)^2l_T^2}{m_T^2}],\\
\frac{1}{m_3^2}&\approx & \frac{1}{m_T^2}[1-\frac{m_c^2-k_T^2}{m_T^2}\frac{\alpha_k^2l_T^2}{m_T^2}].
\end{eqnarray}
Under this approximation, the integrations Eqs.~(\ref{int4}) and (\ref{int5}) will then 
be related to the integrated gluon distribution function $xg(x;Q^2)$,
\begin{eqnarray}
{\cal I}_c\approx \frac{2m_c^2}{36(m_T^2)^3M_X^2}(10M_X^2-27m_T^2)xg(x;m_T^2),\\
{\cal I'}_c\approx \frac{m_c^2-k_t^2}{36(m_T^2)^3M_X^2}(10M_X^2-27m_T^2)xg(x;m_T^2).
\end{eqnarray}
Substituting the above results into Eq.~(\ref{cam}), we can then reproduce the leading
logarithmic approximation result for the diffractive charm jet production process
at hadron colliders which has been calculated in\cite{charm}.

\section{Numerical results}

Provided with the cross section formula (\ref{xsp}) for the partonic process
$gp\rightarrow q\bar q p$, we can calculate the cross
section of the diffractive light quark jet production
at the hadron level.
However, as mentioned above, there exists nonfactorization effect caused by
the spectator interactions in the hard 
diffractive processes in hadron collisions.
Here, we use a suppression factor ${\cal F}_S$ to describe this
nonfactorization effect in the hard diffractive processes at hadron
colliders\cite{soper}.
At the Tevatron,
the value of ${\cal F}_S$ may be as small as ${\cal F}_S\approx 0.1$\cite{soper,tev}.
That is to say, the total cross section of the diffractive processes
at the Tevatron may be reduced down by an order of magnitude due to
this nonfactorization effect.
In the following numerical calculations, we adopt this suppression factor value
to evaluate the diffractive production rate of light quark jet at the Fermilab Tevatron.

The numerical results of the diffractive light quark jet production
at the Fermilab Tevatron are plotted in Fig.3 and Fig.4.
In our calculations, the scales for the parton distribution functions and
the running coupling constant
are both set to be $Q^2=k_T^2$. For the parton distribution functions,
we choose the GRV NLO set \cite{grv}.
In Fig.3, we plot the differential cross section $d\sigma/dt|_{t=0}$
as a function of the lower bound of the transverse momentum of the light
quark jet, $k_{T{\rm min}}$. This figure shows that the cross section is
sensitive to the transverse momentum cut $k_{T{\rm min}}$.
The differential cross section decreases over four orders of magnitude
as $k_{T{\rm min}}$ increases from $5~GeV$ to $15~GeV$.

It is interesting to compare the cross section of the diffractive
light quark jet production with that of the diffractive charm quark jet
production\cite{charm}, which
is also shown in Fig.3. The two curves in this figure show that the production
rates of the light quark jet and the charm quark jet
in the diffractive processes are in the same order of
magnitude.
However, we know that the cross section of diffractive heavy quark jet production
is related to the integrated gluon distribution function, while the cross section
of the light quark jet production is related to the differential
gluon distribution function.

In Fig.4, we plot the differential cross section $d\sigma/dt|_{t=0}$ as
a function of the lower bound of the momentum fraction of the proton
carried by the incident gluon $x_{1{\rm min}}$,
where we set $k_{T{\rm min}}=5~GeV$. This figure shows that the dominant
contribution comes from the region of $x_1\sim 10^{-2}-10^{-1}$.
This property is the same as that of the diffractive charm jet production
at the Tevatron\cite{charm}.

\section{Conclusions}

In this paper, we have calculated the diffractive light quark jet production
at hadron colliders in perturbative QCD by using the two-gluon exchange model.
We find that the production cross section is related to the squared of the
differential gluon distribution function $\partial G(x;Q^2)/\partial ln Q^2$
at the scale of $Q^2\sim k_T^2$, where $k_T$ is the transverse momentum of
the final state quark jet.
For a crossing check, we also used the helicity amplitude method to calculate the diffractive
charm jet production process at hadron colliders, by which we could reproduce the leading
logarithmic approximation result for this process previously calculated in Ref.\cite{charm}.
We have also compared the production rate of the light quark jet in the
diffractive processes with that of the heavy quark jet production,
and found that the production rates of these two processes
are in the same order of magnitude.

As we know, the large transverse momentum dijet production in the diffractive
processes at hadron colliders is important to study the diffractive mechanism and
the nature of the Pomeron. The CDF collaboration at the Fermilab Tevatron
have reported some results on this process\cite{tev}.
In this paper, under the two-gluon exchange model
in perturbative QCD we have calculated the light quark jet production in the diffractive
processes.
In a forthcoming paper, we will calculate the diffractive gluon jet production
in hadron collisions by using the two-gluon exchange model, which will
complete the calculations of the diffractive
dijet production in perturbative QCD.
These results then can be used to compare 
with the experimental measurements to test the valid of the perturbative
QCD description of the diffractive processes at hadron colliders and may be
used to study the diffractive mechanism and the factorization broken effects.

\acknowledgments
This work was supported in part by the National Natural Science Foundation
of China, the State Education Commission of China, and the State
Commission of Science and Technology of China.


\newpage
\vskip 10mm
\centerline{\bf \large Figure Captions}
\vskip 1cm
\noindent
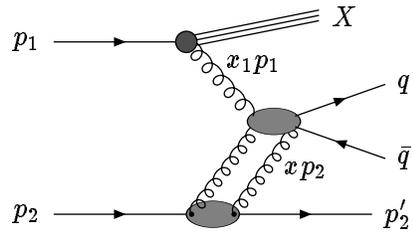
Fig.1. Sketch diagram for the diffractive quark jet production at hadron colliders
in perturbative QCD. 

\noindent
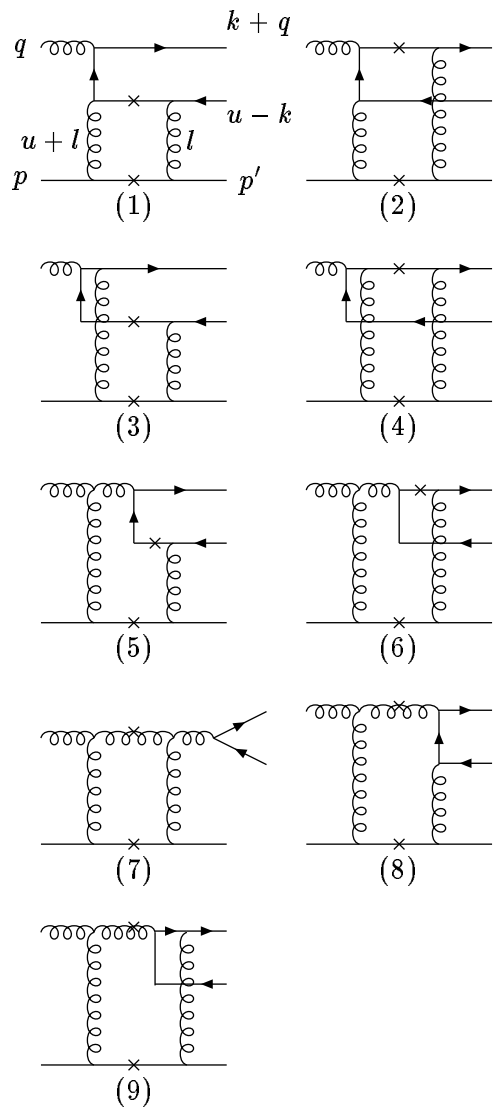
Fig.2. The lowest order perturbative QCD diagrams for partonic process
$gp\rightarrow q\bar q p$.

\noindent
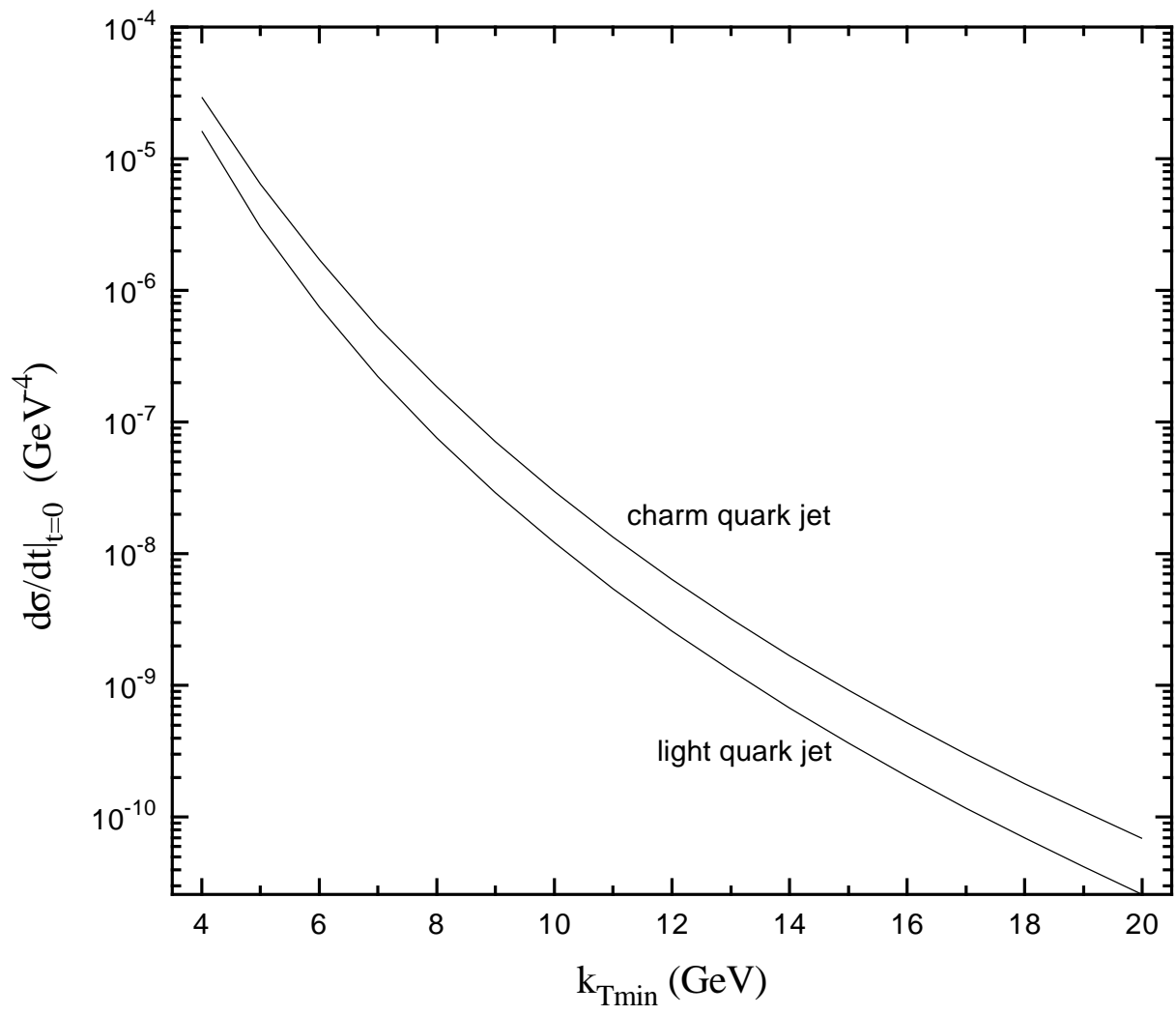
Fig.3. The differential cross section $d\sigma/dt|_{t=0}$ for the light quark jet
production in the diffractive processes as a function
of $k_{T{\rm min}}$ at the Fermilab Tevatron,
where $k_{T{\rm min}}$ is the lower bound of the transverse momentum of the
out going quark jet.
For the charm quark jet production cross section formula, we take from Ref.\cite{charm}
and set $m_c=1.5~GeV$.

\noindent
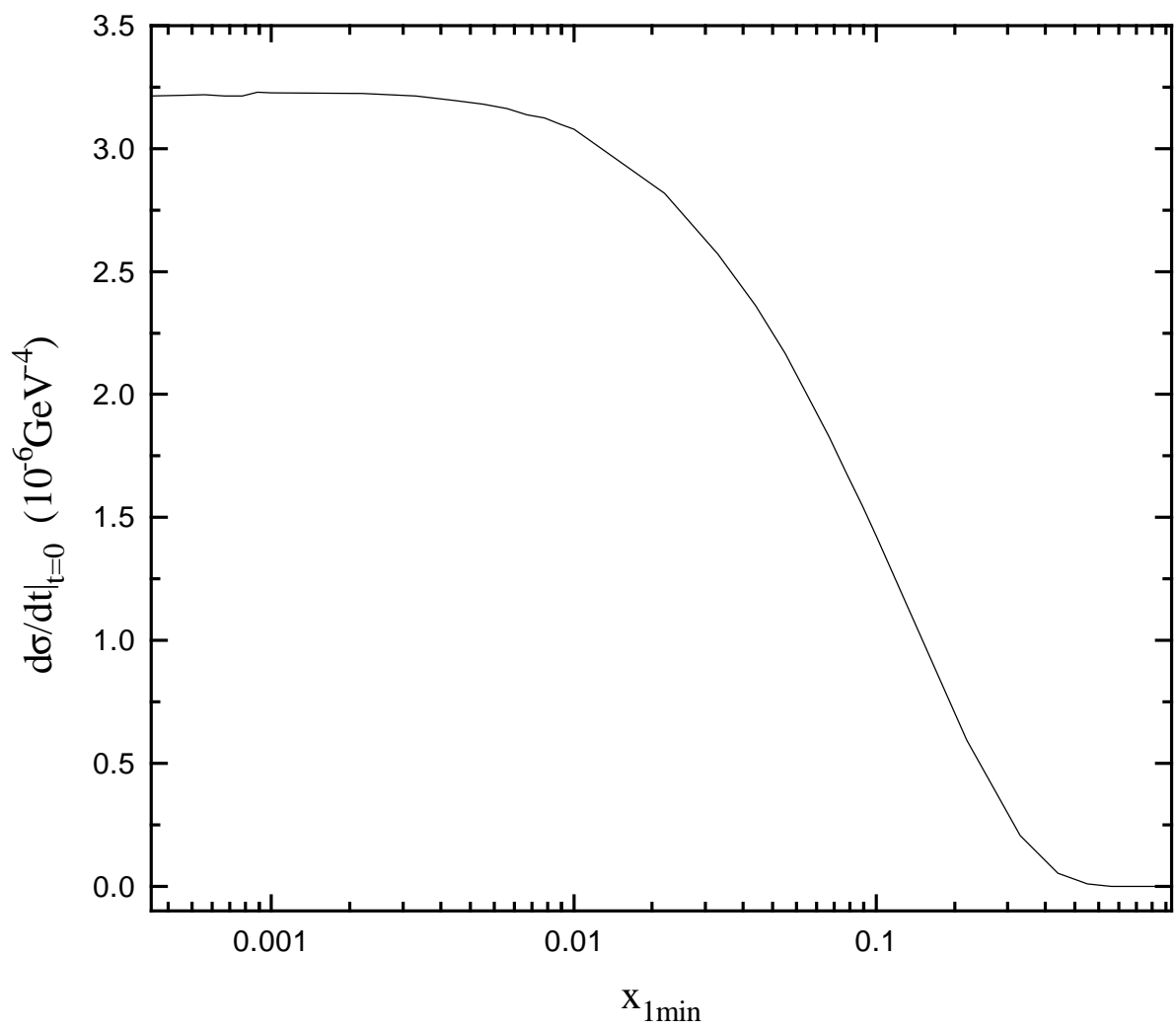
Fig.4. The differential cross section $d\sigma/dt|_{t=0}$ for the light quark jet
production 
as a function of $x_{1{\rm min}}$, where $x_{1{\rm min}}$ is the lower
bound of $x_1$ in the integration of the cross section.

\end{document}